\documentclass[6pt]{elsart}

\usepackage{graphicx}
\graphicspath{{./images/}}
\usepackage{cite}
\usepackage{amsmath}
\usepackage{amssymb}
\usepackage[latin1]{inputenc}

\begin{document}
\begin{frontmatter}
\title{Predictions of highest Transition-temperature for electron-phonon superconductors}
\author{Wei Fan \corauthref{cor1} \thanksref{label1}}
\corauth[cor1]{Corresponding Author : Wei Fan }
\thanks[label1]{Tel : 0086-0551-5591-464; Fax : 0086-0551-5591-434}
\address{Key Laboratory of Materials Physics, Institute of Solid State Physics, Chinese Academy of Sciences, 230031-Hefei, People's Republic of China}
\ead{fan@theory.issp.ac.cn}
\date{\today}
\begin{abstract}
Using the Eliashberg strong coupling theory with vertex correction, we calculate maps of transition temperatures ( T$_{c}$ ) of electron-phonon superconductors in full parameter space. The maximums of transition temperatures for superconductors are predicted based on the maps and the criterion of instability of superconductivity. The strong vertex correction and high transition temperature are tightly correlated in superconductors. We predict that the maximum of T$_{c}$ of new iron-based superconductors will be close to 90(K).
\end{abstract}

\end{frontmatter}

\section{Introductions}

The researches to find home-temperature superconductors are still highlight in
the field of material science~\cite{HomeTc}. The theoretical predictions of
T$_{c}$ are the main efforts of theoretical scientists working in this field.
The BCS theory\cite{Bardeen1} and Eliashberg strong coupling
theory~\cite{Eliashberg1,Nambu1,Allen1,Scalapino1,McMillan1} are well known
acceptable theories to explain the superconductivity of electron-phonon
superconductors. There exists the well known upper limit of T$_{c}$ defined by
the McMillan T$_{c}$ formula~\cite{McMillan1}.  The lattice instability
(softening phonon) and the total strength of electron-phonon coupling also set
the bounds of T$_{c}$ of electron-phonon superconductors~\cite{Moussa1}. The
dielectric properties have been expected to have significant influence on the
possible maximum of T$_{c}$ for a variety of superconductors, but the problem
is still not answered because the complex relation between effective
electron-electron interaction and dielectric function~\cite{Dolgov1}. The
magnetic fluctuation and the vertex correction (non-adiabatic effects)
~\cite{Kostur1} can suppress the T$_{c}$ and lead to the instability of
superconductivity of electron-phonon superconductors. .

The new novel Iron-based high-temperature
superconductor~\cite{Kamihara1,Chen1,Ren1,Wang_C,Cheng_P} has been discovered
and the new record of transition temperature T$_{c}$ has reached about 57.4 (K)
by doping rare earth elements into CaFeAsF~\cite{Cheng_P}. Especially the
significant normal isotope effects of iron with $\alpha\sim 0.4$~ for
Ba$_{1-x}$K$_{x}$Fe$_{2}$As$_{2}$ indicates that electron-phonon interaction
plays an essential role to explain their superconductivity~\cite{LiuRH1}. The
Eliashberg theory suitable for both weak-coupling and strong-coupling should be
studied in more detailed. It is very desirable to study the effects of vertex
correction beyond Migdal theorem~\cite{Kostur1}, especially for superconductors
have components of light atoms. The Eliashberg theory has been successfully
used to calculate T$_{c}$ of many types of
superconductors~\cite{Carbotte1,Moca1,Holcomb1}. However, only special regions
in parameter space of $\lambda-\Omega_{P}-\mu^{*}$ have been explored, where
$\lambda$ is the parameter of electron-phonon interaction, $\Omega_{P}$ the
frequency of phonon and $\mu^{*}$ the Coulomb pseudo-potential. In this paper,
we present the information ( T$_{c}$ map ) in full parameter space
$\lambda-\Omega_{P}-\mu^{*}$. The T$_{c}$ maps obtained in this paper are very
helpful to analyze the relation between T$_{c}$ and these superconducting
parameters and guide to find the superconductors with higher T$_{c}$.

\section{Theory}

In the paper, we generalize the equation of energy gap in
reference~\cite{Kostur1} by including the Coulomb interaction. The standard
energy-gap equation with the form of imaginary Matsubara's
formulation~\cite{Allen1} and the equation after having considered vertex
corrections are used in our calculations. For isotropic   electron-phonon
interaction,
\begin{equation}
H_{ep}=J\sum_{\alpha,pq}c^{+}_{\alpha,p}c_{\alpha,p+q}(a^{+}_{\alpha,-q}+a_{\alpha,q}),
\end{equation}
\noindent the calculations of vertex corrections are greatly simplified. The
electron-phonon interactions are included in the vertex corrections only by the
functions of electron-phonon interaction $\lambda(n)$ defined below. The
self-energy and Green's function of electron in the Nambu scheme are expressed
as
 \begin{eqnarray} \label{Green}
 \nonumber
 \hat{\Sigma}(k,i\omega_{n})&=&(1-Z(k,i\omega_{n}))i\omega_{n}\hat{\tau}_{0}
 -\phi(k,i\omega_{n})\hat{\tau}_{1}+\chi(k,i\omega_{n})\hat{\tau}_{3},
 \\
 \hat{G}(k,i\omega_{n})&=&\frac{i\omega_{n}
 Z(k,i\omega_{n})\hat{\tau}_{0}-\phi(k,i\omega_{n})\hat{\tau}_{1}
 +\varepsilon_{k}(i\omega_{n})\hat{\tau}_{3}}{(i\omega_{n})^{2}Z^{2}(k,i\omega_{n})
 -\varepsilon^{2}_{k}(i\omega_{n})-\phi^{2}(k,i\omega_{n})},
 \end{eqnarray}
\noindent where $\phi(k,i\omega_{n})$ is the function of energy gap,
$Z(k,i\omega_{n})$ the renormalization function, and
$\varepsilon_{k}(i\omega_{n})=\varepsilon^{0}_{k}(i\omega_{n})+\chi(k,i\omega_{n})$.
In above equations $\hat{\tau}_{1}$ and $\hat{\tau}_{3}$ are two Pauli matrixes
and $\hat{\tau}_{0}$ 2$\times$2 unity matrix.

If the density of state of phonon $F(\nu)=V/(2\pi)^{d}\int d^{d}pB(p,\nu)$ and
its spectral function $B(p,\nu)$ are known, the effective electron-phonon
interaction $\alpha^{2}(\nu)$ is defined by
 \begin{eqnarray}\label{AlphaEQ}
 \alpha^{2}(\nu)F(\nu)=\frac{V}{W}
 \int_{S}\frac{d^{d-1}p}{v_{F}}
 \int_{S'}\frac{d^{d-1}p'}{v_{F'}} B(p-p',\nu)|J^{e}_{p-p'}|^{2},
 \end{eqnarray}
\noindent where $W=(2\pi)^{d}\int_{S}d^{d-1}p/v_{F}$, $v^{F}$ Fermi velocity
and $J^{e}_{p-p'}$ the matrix element of electron-phonon interaction which is
constant $J$ in this work. The first order self-energies of electrons
contributed from electron-phonon interaction and electron-electron interaction
are standard and written as
\begin{eqnarray}
 \hat{\Sigma}^{1}(i\omega_{n})&=&k_{B}T\sum_{n'}\lambda(n-n')
 \int_{-E_{F}}^{E_{F}}d\varepsilon_{k'}
 \hat{\tau}_{3}\hat{G}(\varepsilon_{k'},i\omega_{n'})\hat{\tau}_{3} \\
 \nonumber
 \hat{\Sigma}^{c}(i\omega_{n})&=-&k_{B}T\sum_{n'}\mu^{*}
 \int_{-E_{F}}^{E_{F}}d\varepsilon_{k'}
 \hat{\tau}_{3}\hat{G}^{od}(\varepsilon_{k'},i\omega_{n'})\hat{\tau}_{3}
\end{eqnarray}
\noindent where $\hat{G^{od}}(\varepsilon_{k'},i\omega_{n'})$ is off-diagonal
part of Green's function matrix $\hat{G}(\varepsilon_{k'},i\omega_{n'})$. For
convenience, the kinetic energy $\varepsilon_{k'}$ has to shift
$\varepsilon_{k'}-E_{F}$ and a symmetric band from $-E_{F}$ to $E_{F}$ is
chosen. Additionally we assume $k'$-dependent $\hat{G}(k',i\omega_{n'})$
realized by $\hat{G}(\varepsilon_{k'},i\omega_{n'})$. The electron-phonon
interaction is included in
$\lambda(n)=2\int_{0}^{\infty}d\nu\alpha^{2}F(\nu)\nu/(\nu^{2}+\omega_{n}^{2})$.
The Coulomb pseudo-potential is defined as
$\mu^{*}=\mu_{0}/(1+\mu_{0}\ln(E_{F}/\omega_{0}))$, where $\mu_{0}=N(E_{F})U$,
$N(E_{F})$ the density of state at Fermi energy E$_{F}$, U the Coulomb
interaction between electrons and $\omega_{0}$ characteristic energy of typical
phonon correlated to superconductivity.

 The lowest order vertex correction comes from the second order self-energy that is written as
\begin{eqnarray}\label{SE2A}
 \hat{\Sigma}^{2}(k,i\omega_{n})&=&(k_{B}T)^{2}|J|^{4}\Sigma_{n'n''k'k''}
 D(k-k',i\nu_{n-n'})D(k-k'',i\nu_{n-n''}) \\ \nonumber
 &\times&
 \hat{\tau}_{3}\hat{G}(k',i\omega_{n'})\hat{\tau}_{3}
 \hat{G}(k''',i\omega_{n'''})\hat{\tau}_{3}
 \hat{G}(k'',i\omega_{n''})\hat{\tau}_{3}
\end{eqnarray}
\noindent where $D(k^{*},i\nu_{n^{*}})$ is phonon Green's function with the
definitions of indexes $n^{*}$=$n-n'$ or $n-n''$,  $n'''=n'+n''-n$,
$k^{*}$=$k-k'$ or $k-k''$ and $k'''=k'+k''-k$. The summation $\sum_{k^{*}}$ of
wave vector $k^{*}$ ($k^{*}=k'$ or $k''$) is replaced by $V/(2\pi)^{3}\int
d^{3}k^{*}=V/v_{F}(2\pi)^{3}\int d^{2}k^{*}\int d\varepsilon_{k^{*}}$. We
introduce an integral $(1/2E_{F})\int^{E_{F}}_{-E_{F}}d\varepsilon_{k'''}$ in
Eq.(\ref{SE2A}). So the second order self-energy after having averaged on Fermi
surface can be simplified as
\begin{eqnarray} \label{SE2B}
 \hat{\Sigma}^{2}(i\omega_{n})&=&\frac{(k_{B}T)^{2}}{2E_{F}}\Sigma_{n'n''}
 \lambda(n-n')\lambda(n-n'') \\ \nonumber
 &\times&
 \int^{E_{F}}_{-E_{F}}d\varepsilon_{k'}
 \int^{E_{F}}_{-E_{F}}d\varepsilon_{k''}
 \int^{E_{F}}_{-E_{F}}d\varepsilon_{k'''}\\ \nonumber
 &\times&
 \hat{\tau}_{3}\hat{G}(\varepsilon_{k'},i\omega_{n'})\hat{\tau}_{3}
 \hat{G}(\varepsilon_{k'''},i\omega_{n'''})\hat{\tau}_{3}
 \hat{G}(\varepsilon_{k''},i\omega_{n''})\hat{\tau}_{3}
\end{eqnarray}
\noindent now $\varepsilon_{k'''}$ is an independent variable.  Three integrals
of kinetic energy can independently be done by
\begin{equation}
\int^{E^{*}}_{-E^{*}}d\varepsilon_{k^{*}}\hat{G}(\varepsilon_{k^{*}},i\omega_{n})=-\pi
(is_{n}\tau_{0}+\frac{\Delta_{n}}{|\omega_{n}|}\tau_{1})a_{n}
\end{equation}
\noindent when temperature is very close to T$_{c}$,
$\phi(k,i\omega_{n})\rightarrow 0$, so we ignore the term
$\phi^{2}(k,i\omega_{n})$ in the denominator of Green's function
Eq.(\ref{Green}). Other parameters are defined as
$s_{n}=\omega_{n}/|\omega_{n}|$, $a_{n}=(2/\pi)\arctan(E^{*}/|\omega_{n}|)$ and
$\Delta_{n}=\phi(i\omega_{n})/Z(i\omega_{n})$ the energy-gap parameter. In the
general situation, the second-order self-energy cannot simply be written as the
functions of $\lambda(n)$. In terms of above approximation, we transform the
formulas of second-order self energy to a new formation which is suitable for
numerical calculations. Although effects of electron-electron correlation are
not treated accurately, the above approximation is still reasonable because the
main contributions to superconductivity come from $\lambda(n-n')\lambda(n-n'')$
in Eq.(\ref{SE2A}) and Eq.(\ref{SE2B}).

The total self-energy we consider is
$\Sigma(i\omega_{n})=\Sigma^{1}(i\omega_{n})+\Sigma^{c}(i\omega_{n})+\Sigma^{2}(i\omega_{n})$.
By combining with Dyson Equation and considering that, when temperature is very
close to T$_{c}$, $\Delta_{n}\rightarrow 0$, the terms proportional to
$\Delta^{2}$ are ignored and the energy-gap equation is generalized to
\begin{equation} \label{GapEQ}
\sum_{n'=-\infty}^{+\infty}(K_{nn'}-\rho\delta_{nn'})\frac{\Delta_{n'}}{|\omega_{n'}|}=0 , \end{equation}
 \noindent where the kernel matrix
\begin{eqnarray}\label{GapKN}
 K_{nn'}&=&[\lambda(n-n')B(nn')-\mu^{*}+C(nn')]a_{n'}-\delta_{nn'}H_{n'}, \\
 \nonumber
 H_{n'}&=&\sum_{n''=-\infty}^{+\infty}[\delta_{n'n''}
 \frac{|\omega_{n''}|}{\pi k_{B}T}+\lambda(n'-n'')A(n'n'')s_{n'}s_{n''}a_{n''}]
 \end{eqnarray}
\noindent with parameters $A(nn')=1-V_{A}(nn')$, $B(nn')=1-V_{B}(nn')$ and
$a_{n}=(2/\pi)\arctan(E_{F}/Z_{n}|\omega_{n}|)$. The 3-dimensional parameters
of vertex correction have the form
\begin{eqnarray}
 V_{A}(nn')&=&(\pi^{2}k_{B}T/2E_{F})\sum_{n''}\lambda(n-n'')
 s_{n'+n''-n}s_{n''}a_{n'+n''-n}a_{n''} \\ \nonumber
 V_{B}(nn')&=& (\pi^{2}k_{B}T/E_{F})\sum_{n''}\lambda(n-n'')
 s_{n'+n''-n}s_{n''}a_{n'+n''-n}a_{n''}\\ \nonumber
 C(nn')&=&(\pi^{2}k_{B}T/2E_{F})\sum_{n''}\lambda(n-n'')\lambda(n'-n'')
 s_{n'-n''+n}s_{n''}a_{n'-n''+n}a_{n''}.
\end{eqnarray}

If we ignore the vertex corrections, three parameters $V_{A}(nn')$,
$V_{B}(nn')$ and $C(nn')$ are all equal to zero and the kernel Eq.(\ref{GapKN})
of energy-gap equation reduces to the general form without vertex
correction~\cite{Allen1}. At the same time, the differences of the equations of
energy gap in 3-dimensional and 2-dimensional space are only included in the
function $\alpha^{2}F(\nu)$ if vertex corrections are ignored. The introduction
of pair-breaking parameter $\rho$ creates an eigenvalue problem. The physical
gap-equation is corresponding to $\rho$=0. In the calculation of $a_{n}$, we
choose $Z_{n}\sim$1 the value of normal state. The T$_{c}$ is defined as the
temperature when the maximum of eigenvalues E$^{max}$ of kernel matrix
$K_{nn'}$ crosses zero and changes its sign. We use about N=200 Matsubara's
energies to solve above equation. The method to find T$_{c}$ is similar to the
bisection method used to find the roots of a nonlinear
equation~\cite{Bisection}. The nonlinear equation to find T$_{c}$ is
E$^{max}$(T)=0. We can reach the accurate of 0.0001(K) after only 20-30
iterations in temperature interval from 0 (K) to 600 (K).

In calculations of $\alpha^{2}F(\nu)$, we assume that $\alpha^{2}(\nu)$ is
approximately a constant around the peak of phonon mode and the density of
state of phonon is expressed as
\begin{eqnarray}\label{AFEQ}
 F(\nu)=\left\{
 \begin{tabular}{cc}
  $\frac{c}{(\nu-\Omega_{P})^{2}+(\omega_{2})^{2}}
  -\frac{c}{(\omega_{3})^{2}+(\omega_{2})^{2}}$, &
  $|\nu-\Omega_{P}|<\omega_{3}$ \\
  0 & others,
 \end{tabular}
 \right.
 \end{eqnarray}
\noindent where $\Omega_{P}$ is the energy of phonon mode, $\omega_{2}$ the
half-width of peak of phonon mode and $\omega_{3}=2\omega_{2}$. The parameter
of electron-phonon interaction is defined as
$\lambda=\lambda(0)=2\int_{0}^{\infty}d\nu\alpha^{2}F(\nu)/\nu$. An important
relation
 \begin{equation}\label{Meta}
 M\langle\nu^{2}\rangle\lambda=\eta=const
 \end{equation}
\noindent will be used, where $M$ is the effective mass for a certain phonon
mode, especially the constant $\eta$ , the so-called McMillan-Hopfiled
parameter, is closely related to T$_{c}$~\cite{Allen1}. The superconductor
parameter $\eta=N(E_{F})\langle J^{2}\rangle$, characterizes the chemical
environment of atoms and almost keeps as a constant against the simple
structural changes or isotope substitutions.

\begin{figure} \begin{center}\includegraphics[width=0.60\textwidth]{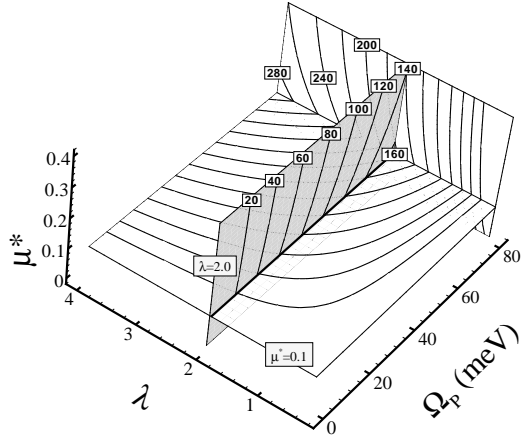}
\caption{\label{fig1} The 3-dimensional T$_{c}$ map and three slice planes with
$\lambda$=2.0, $\mu^{*}$=0.1 and $\Omega_{P}$=80 meV. The numbers within the
rectangles nearby the contour lines are T$_{c}$ values of the corresponding
contour lines.}\end{center}
 \end{figure}

\section{T$_{c}$ Map without vertex corrections}

A T$_{c}$ map in full parameter-space $\lambda-\Omega_{P}-\mu^{*}$ is very
helpful to understand the superconductivity with electron-phonon mechanism. At
first, we choose three-dimensional 30$\times$30$\times$30 mesh-grid with
0.4$\le\lambda\le$4.0, 0$\le\mu^{*}\le$0.4 and 5.0 meV$\le\Omega_{P}\le$80 meV.
We calculate T$_{c}$ on every mesh point using Eq.(\ref{GapEQ}). We plot three
slice planes with $\lambda$=2.0, $\mu^{*}$=0.1 and $\Omega_{P}$=80 meV
respectively shown in Fig.\ref{fig1}. Comparing with others two parameters, the
choice of $\mu^{*}$ has smaller effects on T$_{c}$. The parameter $\lambda$=2.0
is the upper limit of electron-phonon interaction~\cite{McMillan1}. When the
parameter $\lambda$ is larger than 2.0, the crystal lattice will be unstable.
In our calculations, the half-height width $\omega_{2}$ has the same value (4
meV ) for all points of the T$_{c}$ map. If the peak of phonon mode is not too
broad, the T$_{c}$ map has no significant change. Conversely, the Einstein mode
with $\omega_{2}\sim$0 has higher T$_{c}$.

\begin{figure} \begin{center}\includegraphics[width=0.60\textwidth]{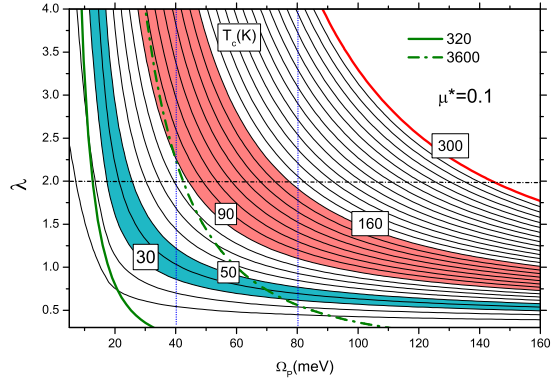}
 \caption{\label{fig2} The T$_{c}$ map in $\lambda-\Omega_{p}$ plane with $\mu^{*}$=0.10
obtained from imaginary-energy Matsubara's method. The bold solid line and
dash-dot line are the curves $\lambda\Omega^{2}_{P}$=320 (meV)$^{2}$
(T$_{c}^{max}$=24 K) and 3600 (meV)$^{2}$ (T$_{c}^{max}$=80 K) respectively.}
 \end{center}\end{figure}

In Fig.\ref{fig2}, we plot a T$_{c}$ map in two-dimensional
$\lambda-\Omega_{P}$ plane in details using 50$\times$100 mesh-grid with
parameters $\lambda$ from 0.4 to 4.0, $\Omega_{P}$ from 5 meV to 160 meV and
the Coulomb pseudo-potential $\mu^{*}$=0.1.  The two-dimensional map is the
extension of the slice of the corresponding 3-dimensional map with
$\mu^{*}$=0.1 shown in Fig.\ref{fig1}. The important role of the relation
$M\langle\nu^{2}\rangle\lambda$=const of Eq.(\ref{Meta}) can be shown clearly
on the T$_{c}$ map. If the effective mass M of a certain phonon mode is
constant, so is the parameter $\langle\nu^{2}\rangle\lambda$. We assume the
half-width of peak of phonon mode $\omega_{2}\ll\Omega_{p}$, so
$\langle\nu^{2}\rangle\lambda \sim \Omega^{2}_{P}\lambda$=const. The two curves
of $\Omega^{2}_{P}\lambda$=320 and 3600 (meV)$^{2}$ are plotted in the same
map. In the regime of strong coupling, $k_{B}T_{c}$ is approximately equal to
$a\sqrt{\langle\nu^{2}\rangle\lambda}$ $\sim$ $a\sqrt{\Omega^{2}_{P}\lambda}$,
where $a$=0.115. Thus $\langle\nu^{2}\rangle\lambda$=const is equivalent to
T$_{c}$=const. In fact, $\langle\nu^{2}\rangle\lambda$=const was used to define
the possible maximum of T$_{c}$~\cite{McMillan1}. The T$_{c}$ increases with
$\lambda$ and saturates at $\lambda=2$ by using the McMillan formula. Similar
to McMillan's idea, our results in the Fig.\ref{fig2} just show that the two
curves coincide approximately with the contour lines of T$_{c}$ in
strong-coupling regime. The approximate saturations happen when $\lambda$ is
very close to 2.

The term $\langle\nu^{2}\rangle\lambda$ characterizes the possible maximum of
T$_{c}$ for fixed $\langle\nu^{2}\rangle\lambda$ value if we consider possible
bounds on $\lambda$.  We can see from the Fig.\ref{fig2} that the curve will
have more chances to meet contour lines with higher T$_{c}$ for larger
$\langle\nu^{2}\rangle\lambda$. Thus, the maximums of T$_{c}$ above two curves
with $\Omega_{P}^{2}\lambda$=320, 3600 (meV)$^{2}$ are close to 20K and 80K
respectively. When superconductor under small structural modifications or
isotope substitution, the corresponding parameters $\lambda$ and $\Omega_{P}$
move along these curves because the chemical environments keep almost
unchanged. In the region of high-energy phonon, the smaller change of $\lambda$
can induce larger change of T$_{c}$ because the curve
$\Omega^{2}_{P}\lambda$=3600 (meV)$^{2}$ spans more contour-lines.

\begin{figure} \begin{center}\includegraphics[width=0.60\textwidth]{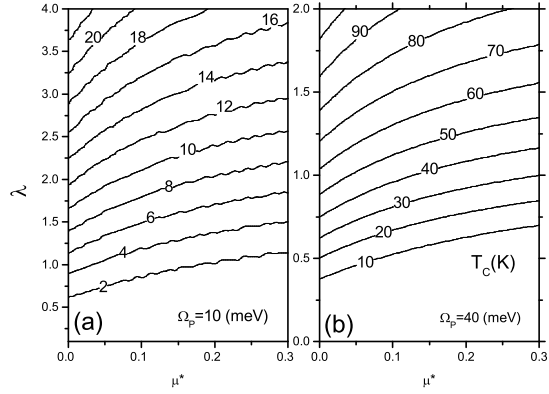}
 \caption{\label{fig3}
 T$_{c}$ maps in $\mu^{*}-\lambda$ plane with $\Omega_{P}$=10 meV (a) and
 40 meV (b). } \end{center}\end{figure}

To study effects of Coulomb pseudo-potential $\mu^{*}$, we calculate two others
T$_{c}$ maps in $\mu^{*}-\lambda$ plane with mesh-grid 50$\times$50 for two
phonon energies $\Omega_{P}$=10 meV and 40 meV respectively (Fig.\ref{fig3}).
For $\Omega_{P}$=10 meV, the maximum of T$_{c}$ is about 24 K, however 100 K
for $\Omega_{P}$=40 meV. We can find that when $\mu^{*}>$ 0.2, $\mu^{*}$ have
smaller influences on T$_{c}$.

\begin{figure} \begin{center}\includegraphics[width=0.60\textwidth]{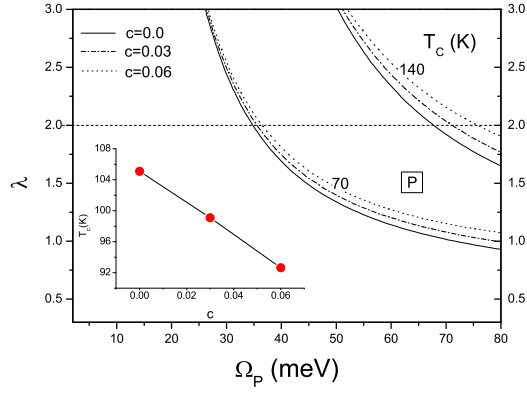}
 \caption{\label{fig4}
The effects of vertex correlations on T$_{c}$ map ($\mu^{*}$=0.10) with two
parameters c=0.12(eV)/E$_{F}$=0.03 and 0.06 compared with the results without
vertex corrections c=0.0 (E$_{F}=\infty$). Two groups of contour lines with
T$_{c}$=70K and 140K are ploted for three values of c=0.0, 0.03 and 0.06
respectively. The inserted figure plots the curve of T$_{c}$ by c at a fixed
point $P$ shown on the $\lambda-\Omega_{P}$ plane. }\end{center}
 \end{figure}

\section{T$_{c}$ Map with vertex corrections}

\begin{figure} \begin{center}\includegraphics[width=0.60\textwidth]{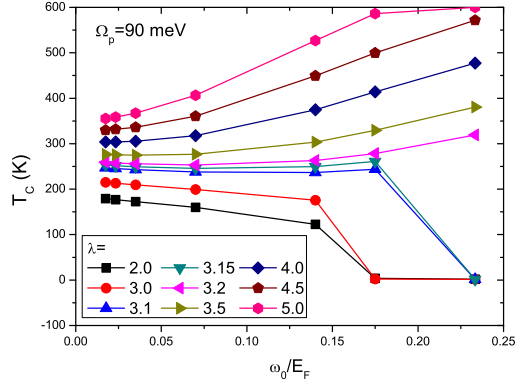}
 \caption{\label{fig5} The positive ($\lambda\geq$3.2) and negative vertex corrections
($\lambda<$3.2) at high phonon energy $\Omega_{P}$=90 meV.}\end{center}
 \end{figure}

It is very important to study the vertex correction in general strong coupling
theory ~\cite{Kostur1}. We control the vertex correction by changing the
Fermi-energy $E_{F}$ in the parameter
$a_{n}=(2/\pi)\arctan(E_{F}/Z_{n}|\omega_{n}|)$. We choose two values 1.0 eV
and 2.0 eV of $E_{F}$.  If E$_{F}$ is too large ($\gg$ 2 eV) the vertex
corrections are invisible in our calculations. Two groups of contour lines with
T$_{c}$=70 K and 140 K are plotted, and for each of groups, there are three
contour lines for c=0.0, 0.03 and 0.06 respectively, where c=0.12(eV)/E$_{F}$
is the control parameter of vertex correction, especially c=0 or E$_{F}=\infty$
is corresponding to the absence of vertex corrections.  We can see that for two
T$_{c}$, the vertex corrections are so small that the T$_{c}$ map has no
significant changes. For the same T$_{c}$, the contour lines only move to
regions with larger $\lambda$ slightly. In the inserted figure of
Fig.\ref{fig4}, at a point $P$ with fixed $\lambda$=1.50 and
$\Omega_{P}$=65meV, the vertex corrections decrease T$_{c}$ to smaller value.
However, the negative vertex corrections only happen when $\Omega_{P}<$80 meV
and $\lambda<$4. Beyond the ranges of parameters, such as $\Omega_{P}$=90 meV
and $\lambda\geq$3.2, the positive vertex corrections (T$_{c}$ increases with
$\omega_{0}/E_{f}$) are realized (Fig.\ref{fig5}). It's not surprised that, if
$\lambda<$3.2, only the general negative vertex-corrections are found although
the phonon energy $\Omega_{P}$ is larger that 80 meV (Fig.\ref{fig5}). The key
point is that, although the effects of vertex corrections are rather
complicated, they do not change the T$_{c}$ maps significantly if E$_{F}$ isn't
too small.

We expect that the vertex corrections are strong for superconductors with high
T$_{c}$. From the Fig.\ref{fig4}, we can see that, when $\Omega_{P}$ is smaller
than 40 meV, the vertex effect is very small although the parameter $\lambda$
of electron-phonon interaction is probably large. The effect of vertex
correction is determined by $\langle\nu^{2}\rangle\lambda$ and not solely by
the parameter $\lambda$ of electron-phonon interaction in the region of low
phonon-energy. The vertex correction is closely correlated to T$_{c}$ because
of $T_{c}\propto\sqrt{\langle\nu^{2}\rangle\lambda}$ in strong coupling regime
with $\lambda >$ 2.0. We will prove that the vertex correction is also
characterized by $\langle\nu^{2}\rangle\lambda$ in the region of low phonon
energy. It easily sees that the parameter
$V_{A}(nn')=P_{V}\sum_{m}\lambda(\nu_{m})s_{n'-m}s_{n-m}a_{n'-m}a_{n-m}$ where
$P_{V}=\pi^{2}k_{B}T_{c}/2E_{F}$ and $m=n-n''$. If the Einstein spectrum
$\alpha^{2}F(\nu)=(\lambda/2)\Omega_{P}\delta(\nu-\Omega_{P})$ is considered,
we can get
\begin{equation}
\label{vertex} \lambda(\nu_{m})=
\lambda\Omega_{P}^{2}/(\nu_{m}^{2}+\Omega_{P}^{2}).
\end{equation}

If $\Omega_{P} \rightarrow 0$, then $\lambda(\nu_{m})\rightarrow
(1/\nu_{m}^{2})\lambda\Omega_{P}^{2}$, and the parameter of vertex correction
$V(nn')\rightarrow P_{V}C(n,n',T_{c})\lambda\Omega_{P}^{2}$, where
$C(n,n',T_{c})=\sum_{m}s_{n'-m}s_{n-m}a_{n'-m}a_{n-m}/\nu_{m}^{2}$ is well
defined and finite. So the parameter $\Omega_{P}^{2}\lambda$ or
$\langle\nu^{2}\rangle\lambda$ characterizes the vertex correction. In real
materials $\Omega_{P}$($\ll E_{F}$) has an up limit $\Omega_{D}$. When
$\Omega_{P}$ approaching to high energy,
$\lambda(\nu_{m})\rightarrow\lambda\Omega_{D}^{2}/(\nu_{m}^{2}+\Omega_{D}^{2})$,
$V(nn')\rightarrow P_{V}D(n,n',T_{c})\lambda$, where
$D(n,n',T_{c})=\sum_{m}s_{n'-m}s_{n-m}a_{n'-m}a_{n-m}
\Omega_{D}^{2}/(\nu_{m}^{2}+\Omega_{D}^{2})$ is well defined. So we can see
that the vertex correction is characterized by $\lambda$ at high energy. From
the Fig.\ref{fig2}, T$_{c}$ is almost independent on $\Omega_{P}$ when
$\Omega_{P}>$ 40 meV and $\lambda<2.0$, and T$_{c}$ is determined by $\lambda$
as well. Thus, we have proven that the vertex correction (as well as T$_{c}$)
is characterized by $\langle\nu^{2}\rangle\lambda$ in the region of low phonon
energy and by $\lambda$ in the region of high phonon energy. There is a
characteristic energy close to 40 meV and the vertex corrections (as well as
T$_{c}$) have different behavior when $\Omega_{P}$ is larger or smaller than
the characteristic energy.

\section{Discussion and Summary}

For superconductors with T$_{c}$ close to home temperature, the parameter
$\lambda$ of electron-phonon interaction should be larger than 2.0 or the
phonon-energy $\Omega_{P}$ larger than 80 meV (Fig.~\ref{fig2}). Generally, we
hope to find higher T$_{c}$ by increasing the frequency of phonon by choosing
compounds including light atoms such as Hydrogen atoms in molecule
crystals~\cite{Eremets1}. The contour line T$_{c}$=300K in Fig.\ref{fig2} is
beneath the line of $\lambda=$2.0 when $\Omega_{P}$ is larger than 150 meV or
1210 cm$^{-1}$ for Raman shift. We can see that only small changes of $\lambda$
can induce large changes of T$_{c}$ if $\Omega_{P}>$ 80 meV. In compounds with
components of light atoms and heavy atoms, the frequencies of optical modes are
determined by light atoms. The heavy atoms have the capability to stabilize the
lattice vibrations so it is advantaged to form macroscopic-coherent
superconducting states.

It is very interesting that we can explain the spatial anti-correlation between
energy-gap and phonon energy of cuprate superconductor Bi2212~\cite{Lee1} in
terms of the electron-phonon mechanism by having solved the real-energy
Eliashberg Equation under the constraint from Eq.(\ref{Meta})~\cite{Fan1}. The
large value of $\lambda$ ($>$1.0) can be obtained from the frozen-phonon method
by considering nonlocal long-range Madelung interaction which is non-adiabatic
effect when an atom displaces away from its equilibrium position
~\cite{Jarlborg1,Krakauer1}. Our results in this context also indicate that the
vertex corrections which include part of non-adiabatic effect are closely
correlated to high T$_{c}$. On the other hand, in terms of the linear
temperature-dependent resistivity at high temperature above T$_{c}$, the strong
electron-phonon coupling could be extracted by the analysis of strong coupling
theory~\cite{Zeyher1,Mazin1}. The upper limit of phonon energies of in-plane
breathing modes of cuprate superconductors is close to $\Omega_{P}$=80
meV~\cite{Zhao1}. If the anisotropy of Bi2212 crystal is ignored, from the
Fig.\ref{fig1}, we can see that the maximum of T$_{c}$ will be within
140K-160K. The highest T$_{c}$ is about 130 K for cuprate superconductors close
to the range. From the Fig.\ref{fig4}, there are larger vertex corrections at
higher $\Omega_{P}$ and larger $\lambda$, however the T$_{c}$ doesn't changes
significantly because the band-width of CuO$_{2}$ plane for cuprate
superconductors such as Bi2212 is not too small about 4 eV, and the
Fermi-energy $E_{F}$ is at least 1 eV~\cite{Krakauer2,Lin1}.

Finally, we discuss the new iron-based layered high-temperature superconductors
~\cite{Kamihara1,Chen1,Ren1,Wang_C,Cheng_P}.  The carriers are mostly possible
in FeAs layers.  The multi-band electronic structure, especially the
cooper-pairs spanning different Fermi-surface sheets probably play an important
role to its superconductivity. The pairing mechanism of different Fermi-surface
sheets is absent in present model, but the model is still equivalent to the
mean of multi-band model. Very similar to the cuprate superconductors, the
linear-response calculation of electron-phonon interaction only provides very
small parameter $\lambda$ of electron-phonon coupling that does not account the
requirement of high T$_{c}$ from 27(K) to 57(K)~\cite{Boeri1}. However, just
like for cuprate superconductors~\cite{Krakauer1}, the nonlocal long-range
Madelung interaction generally enhances parameter $\lambda$ of electron-phonon
interaction and account the requirements for high T$_{c}$. Experimental
evidence for strong electron-phonon coupling comes from the measurement of
normal-state resistivity at high temperature~\cite{Bhoi1}. The upper limit of
phonon-energy is about 40 meV~\cite{Higashitaniguchi1}. From the Fig.\ref{fig2}
and Fig.\ref{fig3}(b), we can predict that the maximum of transition
temperature T$_{c}$ for layered iron-based superconductors may access to 90(K).

In summary, we calculate T$_{c}$ maps in full parameter space using strong
coupling theory including vertex corrections. We also discuss the possibility
increasing T$_{c}$ by using materials including light atoms with higher phonon
frequency. The strong vertex corrections are inevitable in materials with high
T$_{c}$. Based on the maps obtained in this work, we predict the maximum of
T$_{c}$ of new finding iron-based superconductors is about 90(K).

\section{Acknowledgement}

The author gratefully thanks Dr. Li Yang-Ling for reading and criticizing the
manuscript. This work is supported by Director Grants of Hefei Institutes of
Physical Sciences, Knowledge Innovation Program of Chinese Academy of Sciences
and National Science Foundation of China. The eigenvalues of kernel matrix are
obtained by using the Lapack package.

\end{document}